# Deep Learning-based Assessment of the Relation Between the Third Molar and Mandibular Canal on Panoramic Radiographs using Local, Centralized, and Federated Learning

Johan Andreas Balle Rubak[1,][*], Sara Haghighat[1], Sanyam Jain[1], Mostafa Aldesoki[2], Akhilanand Chaurasia[3], Sarah Sadat Ehsani[4], Faezeh Dehghan Ghanatkaman[5], Ahmad Badruddin Ghazali[6], Julien Issa[7], Basel Khalil[8], Rishi Ramani[9] and Ruben Pauwels[1]

[1] Department of Dentistry and Oral Health, Aarhus University, Vennelyst Boulevard 9, Aarhus, 8000, Denmark; ruben.pauwels@dent.au.dk
[2] Private Dental Practice Dr. Jörg Deike, Hannover, Germany; mostafadesoki86@gmail.com
[3] Department of Oral Medicine and Radiology, King George's Medical University, India. Shah Mina Road, Chowk, Lucknow, Uttar Pradesh 226003, India; chaurasiaakhilanand49@gmail.com
[4] Department of Diagnosis and Oral Health, University of Louisville School of Dentistry, Louisville, KY , 40202, USA; Sarah.esn20@gmail.com
[5] Faculty of Dentistry, Islamic Azad University of Medical Sciences, Tehran, Iran; faezedehghan963@gmail.com
[6] Department of Oral Maxillofacial Surgery & Oral Diagnosis, Kulliyyah of Dentistry, International Islamic University Malaysia; badruddinghazali@iium.edu.my
[7] Department of Oral Radiology & Digital Dentistry, Academic Centre for Dentistry Amsterdam (ACTA), University of Amsterdam and Vrije Universiteit Amsterdam, Gustav Mahlerlaan 3004, 1081 Amsterdam, The Netherlands; j.i.isaa@acta.nl
[8] Private Practice, Stockholm, Sweden; baselkhalil51@outlook.com
[9] Melbourne Dental School, University of Melbourne, Victoria, Australia; r.ramani@unimelb.edu.au
[*] Correspondence: johan.a.rubak@gmail.com

**Featured Application**

This framework enables privacy-preserving, multi-center development of automated tools to triage mandibular third molar cases on panoramic radiographs and identify patients at increased risk of inferior alveolar nerve involvement.

**Abstract**

Impaction of the mandibular third molar in proximity to the mandibular canal increases the risk of inferior alveolar nerve injury. Panoramic radiography is routinely used to assess this relationship. Automated classification of molar–canal overlap could support clinical triage and reduce unnecessary CBCT referrals, while federated learning (FL) enables multi-center collaboration without sharing patient data. We compared Local Learning (LL), FL, and Centralized Learning (CL) for binary overlap/no-overlap classification on cropped panoramic radiographs partitioned across eight independent labelers. A pretrained ResNet-34 was trained under each paradigm and evaluated using per-client metrics with locally optimized thresholds and pooled test performance with a global threshold. Performance was assessed using area under the receiver operating characteristic curve (AUC) and threshold-based metrics, alongside training dynamics, Grad-CAM visualizations, and server-side aggregate monitoring signals. On the test set, CL achieved the highest performance (AUC 0.831; accuracy ≈ 0.782), FL showed intermediate performance (AUC 0.757; accuracy ≈ 0.703), and LL generalized poorly across clients (AUC range ≈ 0.619–0.734; mean ≈ 0.672). Training curves suggested overfitting, particularly in LL models, and Grad-CAM indicated more anatomically focused attention in CL and FL. Overall, centralized training provided the strongest performance, while FL offers a privacy-preserving alternative that outperforms LL.

## 1. Introduction

Impaction of the mandibular third molar in close proximity to the mandibular canal is a common clinical scenario. The prevalence of impacted third molars is reported to be 36.9%, with 38-60% showing contact or close proximity to the mandibular canal in another study [1, 2]. This impaction carries a non-negligible risk of inferior alveolar nerve injury during extraction [3, 4, 5]. Preoperative assessment of the relation between the third molar root and the mandibular canal (3M-MC) therefore plays a central role in surgical planning. While cone-beam computed tomography (CBCT) provides excellent three-dimensional detail [6], panoramic radiographs (PR) remain the most widely used initial imaging modality due to lower cost, lower radiation dose, and broad availability; therefore it is necessary to develop reliable automated methods for assessing the 3M-MC relation on PRs, and thereby improve screening and triage, which would reduce unnecessary CBCT referrals, and standardize risk assessment across centers [7, 8, 6].

Recent work has demonstrated that deep learning can assist in 3M-MC assessment on PRs [9], often using convolutional backbones (e.g., ResNet variants) for model training and either radiologist consensus or CBCT-confirmed labels as reference standard to confirm the true anatomical relationship. These prior results show the feasibility and clinical promise of automated 3M-MC assessment, while also highlighting sensitivity of performance to dataset size, label quality, and imaging heterogeneity [10, 7, 11, 12, 13, 14].

A major barrier to improving generalization and robustness of medical imaging AI models is the difficulty of assembling large, diverse, multi-institutional datasets. Privacy, governance, and logistical constraints frequently prevent medical images from being centralized. Federated learning (FL) has therefore emerged as a promising alternative. FL enables collaborative model training across multiple sites (from this referred to as 'clients') without exchanging the underlying data, instead aggregating model updates or gradients [15]. FL has attracted substantial attention in the medical imaging community and has been applied to multiple tasks across various modalities [16, 17, 18].

Despite its appeal, FL faces important practical challenges in real-world medical deployments [19]. Most concerning is data heterogeneity, i.e., non-Independent and Identically distributed (IID) client distributions. Clients differ in population case-mix, imaging devices and protocols. Furthermore, crucially for annotation-driven tasks in supervised learning, labeling practices and class distribution can vary between clients. Pathological or clinically important classes are often rare relative to normal instances, which biases standard training procedures toward the majority class and inflates commonly reported accuracy metrics. This also complicates this federated aggregation and necessitates post-processing strategies to mitigate imbalance [20, 21, 22].

Empirical studies show that heterogeneity can substantially degrade FL performance relative to centralized training, and that different forms of heterogeneity (label distribution skew, feature distribution differences, and quantity imbalances) have distinct effects on model convergence and final accuracy [23]. Annotation variability across labelers (inter-labeler disagreement) further complicates federated aggregation, because labeler-specific biases become entangled with client-specific data distributions [24]. A particular issue in FL is that such issues are much more difficult to detect during a training process due to data opacity [19], where judgments regarding the data and annotation quality of a client have to be based on model weights and gradients [22, 23, 24].

In this study, we perform a focused **empirical** comparison of three learning paradigms for binary classification of 3M-MC relation on PR: **(1) Local Learning (LL)** — independent models trained only on each client's own data; **(2) Centralized Learning (CL)** — a single model trained on pooled data from all clients, and **(3) Federated Learning (FL)** — a collaborative model obtained by federated aggregation across the same clients, without centralizing data.

The per-labeler partitioning of data in our experimental design allows us to directly isolate the effects of data quantity, labeler-specific bias, and class prevalence on each training paradigm. Our knowledge contributions therefore include (i) the influence of client-level class imbalance and simple imbalance-correction strategies; (ii) the sensitivity of FL aggregation to heterogeneous label distributions; and (iii) the role of annotation variability in producing a federated performance gap. These analyses shed light on the performance of federated models in a realistic federated setup in comparison to Local and Centralized Learning. Our hypothesis is that even in a non-IID scenario, FL is superior to LL and non-inferior to CL.

## 2. Materials and Methods

### 2.1. Data

The dataset utilized in this study comprised panoramic radiographs sourced from five publicly available datasets: ADLD [25], Dentex [26], TSXK [27], Tufts [28], and USPFORP [29]. An automated process was employed to select a region of interest (ROI), specifically the lower third molars, on each PR. The process used the tooth annotations provided by each dataset; based on these annotations, an expanded bounding box was defined with the third molar at the center.

A detailed protocol was established to guide the labeling process for a team of eight professionally trained dentists with average clinical experience of seven years. To ensure consistency a subsample of 50 images was provided to all labelers for computing pairwise inter-observer reliability using quadratic-weighted Cohen's κ. Subsequently, all images were distributed among the labelers (clients) for class assessment based on the 3M-MC relationship, based on the following scheme:

1. Root-canal relation
    a. Root complex entirely superior to the canal
    b. Root complex overlapping the canal borders
    c. At least part of the root complex inferior to the canal
2. Location of overlap (if present)
    a. Apical part in overlap
    b. Middle part in overlap
    c. Cervical (coronal) part in overlap

Because only a limited number of overlap cases were available, all overlap subclasses (2A–2C, 3A–3C) were merged into a single overlap class, while class 1 remained as no-overlap, yielding a binary classification task. Table 1 shows the class imbalance and client-level label skew. The final number of examples per class for each client and for the centrally pooled dataset is also reported.

*2.2. Model*

A ResNet-34 convolutional backbone was used as the reference model for all experiments. ResNet-34 is a widely used, well-tested residual network that provides a useful balance between representational capacity and computational cost. Choosing a standard, established backbone as a baseline simplifies comparison to previous work and reduces confounding variables during iterative development [30, 11]. The network was initialized with ImageNet pretrained weights, and the final classification head was replaced with a two-unit linear layer for binary classification (no-overlap vs. overlap). Input images were adapted to the network input size and normalized with ImageNet mean/std prior to training.

**Table 1.** Per-client and pooled counts for all original classes and pooled binary counts (no-contact/contact).

| Client | Dataset | 1 | 2A | 2B | 2C | 3A | 3B | 3C | No-overlap | Overlap | Total | Overlap % |
|---|---|---|---|---|---|---|---|---|---|---|---|---|
| Client 0 | adld | 570 | 58 | 11 | 7 | 0 | 3 | 1 | 570 | 80 | 650 | 12.3% |
| Client 1 | adld | 386 | 238 | 4 | 21 | 0 | 0 | 1 | 386 | 264 | 650 | 40.6% |
| Client 2 | adld | 397 | 199 | 8 | 4 | 9 | 1 | 1 | 397 | 222 | 619 | 35.9% |
| Client 3 | adld | 388 | 196 | 29 | 38 | 0 | 0 | 0 | 388 | 263 | 651 | 40.4% |
| Client 4 | dentex | 484 | 230 | 18 | 38 | 9 | 13 | 1 | 484 | 309 | 793 | 39.0% |
| Client 5 | tsxk | 401 | 167 | 8 | 13 | 13 | 8 | 3 | 401 | 212 | 613 | 34.6% |
| Client 6 | tufts | 499 | 173 | 21 | 40 | 21 | 4 | 1 | 499 | 260 | 759 | 34.3% |
| Client 7 | uspforp | 436 | 208 | 3 | 5 | 10 | 15 | 4 | 436 | 245 | 681 | 36.0% |
| Pooled | N/A | 3561 | 1469 | 102 | 166 | 62 | 44 | 12 | 3561 | 1855 | 5416 | 34.3% |

*2.3. Learning Paradigms*

The three paradigms (illustrated in Figure 1) differ in how data and model updates are shared:

- **Local Learning (LL).** Each client trains a model using only its own local data and keeps both data and model weights local. LL serves as a lower bound in our comparisons because each model sees only a fraction of the total data and is subject to client-specific biases and class-skew.
- **Centralized Learning (CL).** All clients send their labeled data to a central location where a single model is trained on the pooled dataset. CL typically yields the best performance when pooling is permissible, since the model benefits from maximal data diversity and sample size, but it requires data transfer and raises privacy/governance concerns.
- **Federated Learning (FL).** Clients keep raw data locally and collaboratively train a shared global model by exchanging model parameters or updates. FL aims to approximate the performance of CL while avoiding transfer of sensitive images. FL is well suited to multi-center medical imaging scenarios but is sensitive to non-IID client distributions, communication constraints and aggregation strategy choices [31].

In this work, the clients correspond to the data subsets annotated by the eight independent labelers, as each labeler annotated a separate subset of ROIs and thus defined a natural client partition. Note that the data from the five abovementioned public sources was kept segregated between labelers, as shown in Table 1; in other words, a given labeler only assessed images from a single public dataset, maintaining image quality and population heterogeneity reflecting a real-world situation. All experiments use the same ResNet-34 architecture and the same held-out test set so that differences between paradigms reflect the training paradigm rather than architectural or evaluation differences.

We implemented a standard rounds-based FL workflow. One round *t* proceeds as follows:

1. The server holds the current global weights $w^t$ and distributes them to the participating clients.
2. Each client *k* initializes from $w^t$ and performs local training for *E* epochs on its local dataset, producing updated weights $w_k^{t+1}$.
3. Clients send their updated weights back to the server.
4. The server aggregates the client updates into a new global model. With Federated Averaging (FedAvg) [31, 32] the aggregation is a sample-weighted average:

$$w^{t+1} = \sum_{k=1}^{K} \frac{n_k}{n} w_k^{t+1},$$

where $n_k$ is the number of training examples at client *k* and $n = \sum_k n_k$.

5. Steps 1–4 are repeated for multiple rounds until the global model converges or a stopping condition is met.

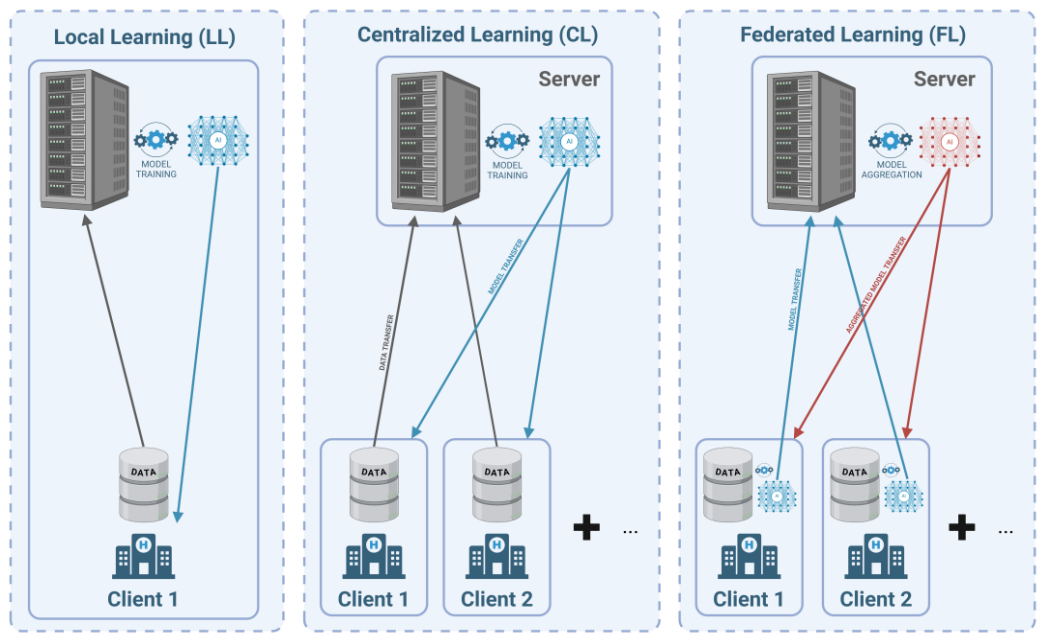

**Figure 2.** Learning Paradigms: to the left Local Learning, in the middle-Centralized Learning and to the right Federated Learning. Created in BioRender. Rubak, J. (2026) https://BioRender.com/sob1383 [29].

Practical FL design choices that impact performance include the number of local epochs E, client selection per round, the aggregation weighting (sample counts vs. uniform), and whether to apply server- or client-side techniques to mitigate heterogeneity (e.g., class-weighted losses, balanced sampling, or optimization variants such as FedProx). In our experiments, we implemented the setup using Flower AI [33] and the canonical FedAvg protocol as the baseline federation strategy.

*2.4. Training Procedure*

The training pipeline was kept identical across LL, CL and FL to ensure a fair and controlled comparison. The data management is illustrated in Figure 2. First, the full labeled dataset was pooled and a class-stratified, centralized test set of 10% was sampled. The remaining 90% of the data was redistributed back to the original clients based on their source assignment. From these per-client splits a random validation subset (with preserving class imbalance) was selected such that the total amount of validation data across clients corresponded to 10% of the original dataset. Because the test set had already been removed, these local validation subsets represent slightly more than 10% of each client's remaining data (approximately 11.1% of the post-test pool). These per-client validation sets were used to monitor training progress in LL and FL, and their union served as the centralized validation set in CL. Although the use of a centrally pooled test set simplifies comparison, it does not reflect privacy-restricted deployment scenarios; therefore, we additionally evaluated final models on each client's local validation/test subset to assess per-client generalization.

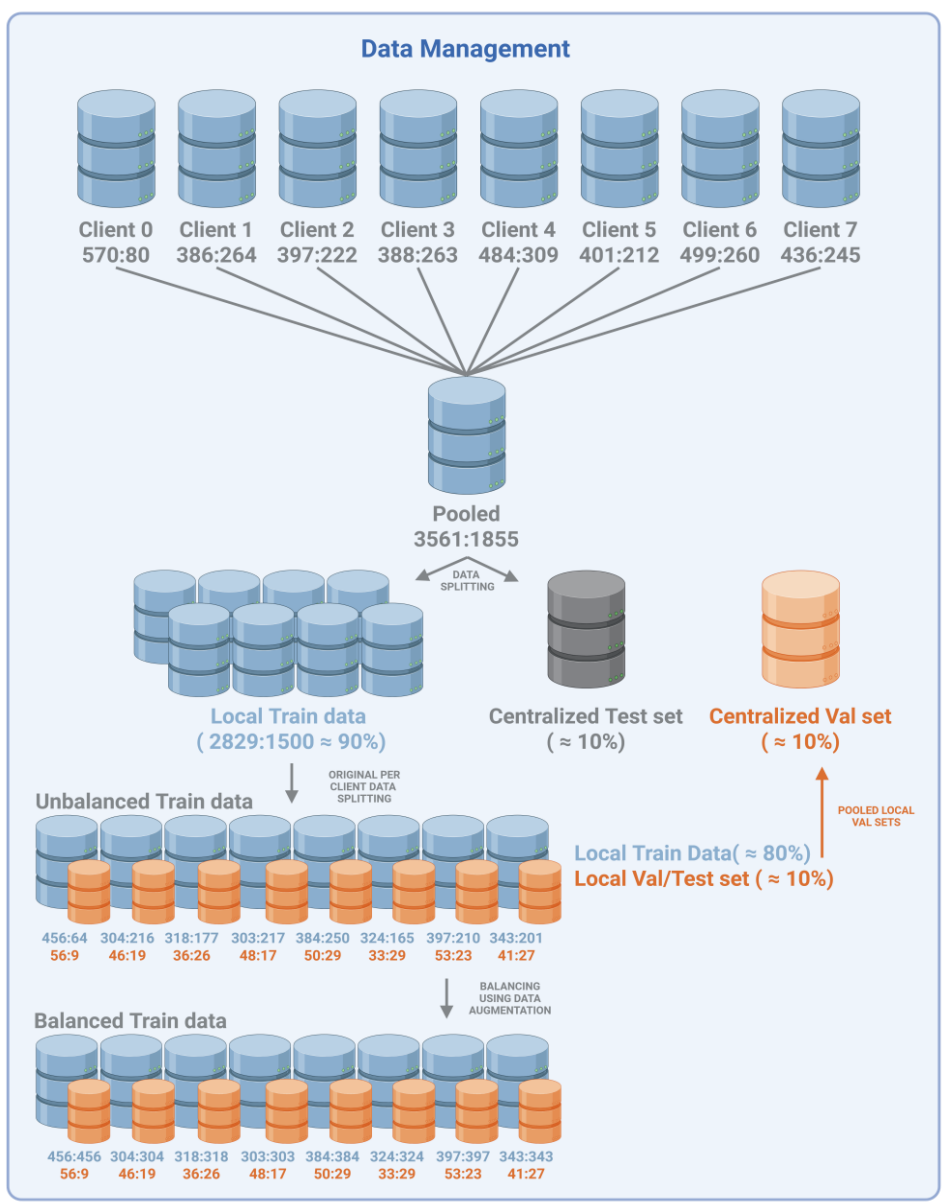

**Figure 2.** Data management, showing initial split of test data, based on data from all participating clients. Then the following partitioning into training and validation sets for each client and final balanced train datasets using extra data augmented examples. Created in BioRender. Rubak, J. (2026) https://BioRender.com/f0dsfs6.

All images passed through the same preprocessing pipeline. Contrast Limited Adaptive Histogram Equalization (CLAHE) [34] was applied to the luminance channel to enhance contrast in panoramic radiographs, then ROIs were resized to 224 x 224, and intensities were normalized using ImageNet's mean and standard deviation. To address class imbalance, the minority class was upsampled by repeatedly picking minority-class images at random (allowing repeats) and applying image augmentations to create extra copies until the two classes are balanced; each augmented copy uses a fixed per-item random seed so it looks the same during an epoch, but can be refreshed when the set is regenerated. This was done after initial train and validation splits were created. The data augmentation applied included reflecting padding of size 16, random horizontal flips, small random rotations (≈ 3°), center cropping after padding, slight color jitter, and stochastic Gaussian blur (kernel size = 5, sigma = 0.5). Importantly, the augmented samples were regenerated periodically: in LL and CL the balanced augmented dataset was recreated every second epoch, while in FL it was naturally regenerated at each round, since clients rebuild their local datasets each time they perform local training. This periodic regeneration introduces new stochastic augmentations over time, improving generalizability.

For all paradigms, the model was trained using a Binary Cross Entropy with Logits loss function, a threshold of 0.5 and batch size of 32. Optimization used the AdamW optimizer with a learning rate of $1 \times 10^{-4}$, weight decay of $1 \times 10^{-5}$. The number of epochs in CL and LL were 10, and in FL 5 rounds of 2 epochs were used to ensure a fair comparison. Training was performed on a NVIDIA RTX 6000 Ada Generation GPU with 48 GB of graphics memory.

Final evaluation was performed on the centralized 10% test set. In addition, all models were evaluated on the per-client validation splits to examine variability and robustness across clients and to better reflect heterogeneous real-world conditions.

*2.6 Performance Metrics and Statistical Analysis*

To evaluate and compare the performance of LL, FL, and CL paradigms, a comprehensive set of quantitative metrics and visualization tools was employed. Training progress was tracked throughout the optimization process for all paradigms by monitoring training loss, validation loss, and validation accuracy across epochs (LL and CL) or communication rounds (FL). This provided insight into convergence behavior, model stability, and potential tendencies toward underfitting or overfitting. For final model evaluation, several standard binary classification metrics were computed. Let TP, FP, TN, and FN denote the number of true positives, false positives, true negatives, and false negatives, respectively. The following metrics were used:

$$Accuracy = \frac{TP + TN}{TP + TN + FP + FN}$$

$$Sensitivity\ (Recall) = \frac{TP}{TP + FN}$$

$$Specificity = \frac{TN}{TN + FP}$$

$$Precision = \frac{TP}{TP + FP}$$

$$F1\ score = 2 * \frac{Precision * Recall}{Precision + Recall}$$

$$Balanced\ Accuracy = \frac{1}{2}(Sensitivity + Specificity)$$

$$J\ (Youden's J) = Sensitivity - (1 - Specificity)$$

In addition, the area under the receiver operating characteristic curve (AUC) was computed as a threshold-independent measure of discriminative ability, defined as

$$AUC = \int_0^1 TPR(FPR)\ d(FPR)$$

where TPR and FPR denote the true positive rate and false positive rate at a given decision threshold.

Visual analytical tools were included to complement the numerical metrics. Gradient weighted Class Activation Mapping (Grad-CAM) visualizations were generated for representative samples to qualitatively assess model interpretability and to verify whether each model focused on anatomically meaningful regions during prediction. Furthermore, each model paradigm was evaluated under two conditions: (i) using optimized thresholds derived individually from each local validation set (LL, CL, and FL evaluated locally), and (ii) using a global threshold optimized on the pooled training and validation data from all clients (evaluation on the centralized test set). This two-level evaluation design allowed assessment of both local-domain performance and the ability of each paradigm to generalize across heterogeneous client distributions.

To assess statistical significance of performance differences between learning paradigms, AUC values were compared using nonparametric and parametric tests as appropriate. Normality of paired AUC differences across clients was first assessed using the Shapiro–Wilk test; however, given the small sample size n=8 and to remain robust to distributional assumptions, paired Wilcoxon signed-rank tests were used for all comparisons on local validation sets. For evaluation on the centralized test set, differences in AUC between models were assessed using DeLong's test [35] for correlated ROC curves. Where multiple related hypotheses were tested within the same evaluation setting. Bonferroni correction was applied to control the family-wise error rate; primary comparisons were evaluated without correction.

## 3. Results

### 3.1. Calibration & Training Dynamics

Pairwise quadratic-weighted Cohen's κ for the craniocaudal label across clients is shown in Appendix A.2, Figure 8. Across the unique observer-pairs, a mean weighted κ = 0.81, median = 0.80, and range = 0.69–0.97 was calculated. Most client–client pairs show substantial agreement (κ ≈ 0.75–0.90); the lowest pairwise values (≈0.69–0.73) involve Client 3, indicating relatively lower agreement compared with others.

All models exhibited rapid reduction in training loss and converged to low training-loss values within the run (note that training for 50 epochs was carried out initially to confirm that convergence was received already withing 10 epochs). Validation loss, however, was more variable and in many cases increased across epochs/rounds while validation accuracy showed a slow, noisy upward trend. This pattern is visible across the LL, FL and CL curves: training loss curves are smooth and convergent, whereas validation loss curves fluctuate and occasionally rise between epochs/aggregation rounds (Figure 3, left/center panels). Validation accuracies (Figure 3, right panels) increase on average but remain substantially more variable than the training curves; the fluctuation is particularly evident for FL after each aggregation round.

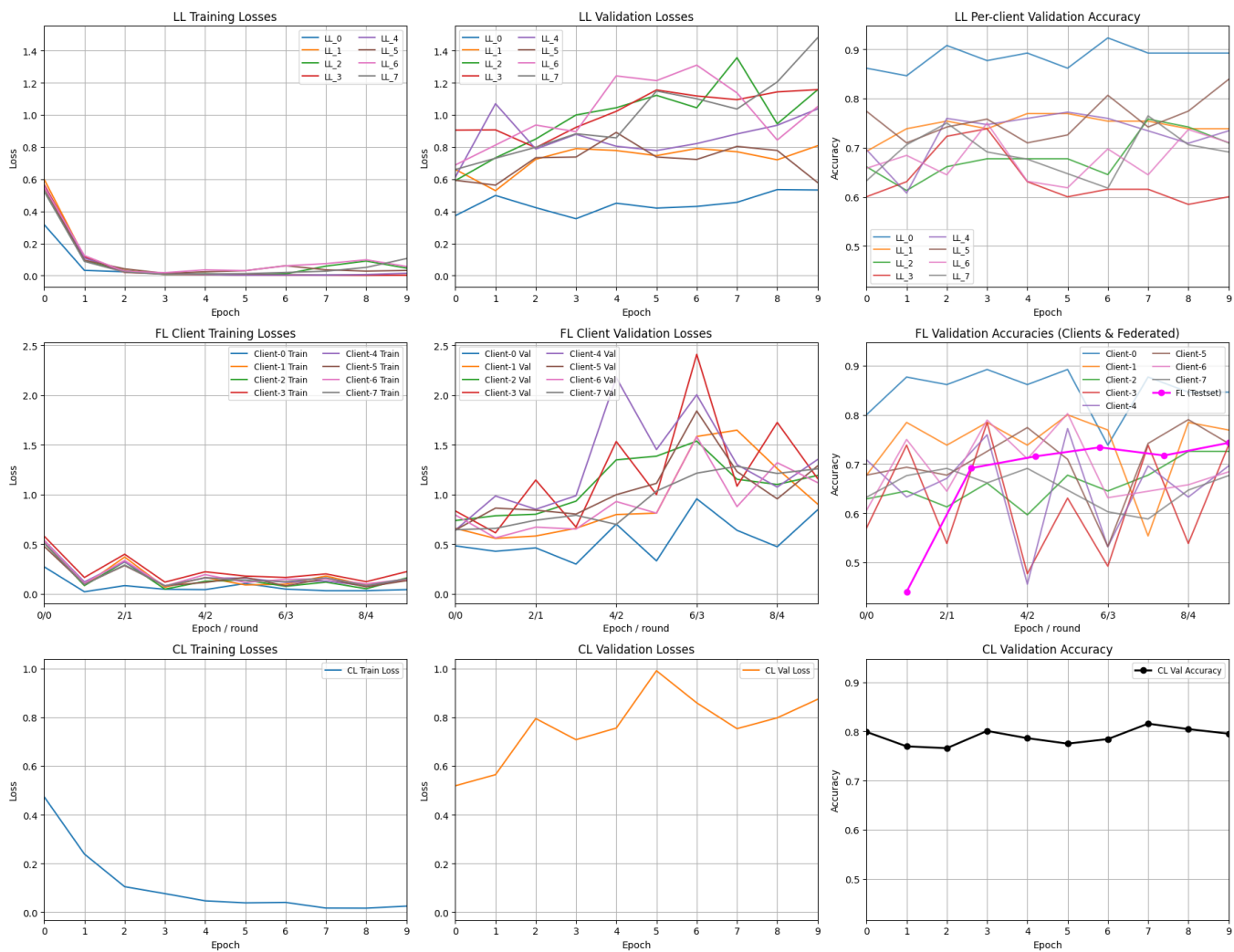

**Figure 3.** Training and validation behavior for three learning setups (LL = local client training, FL = federated learning, CL = centralized/pooled training). Top row — LL (per-client) results: left, per-client training loss over 10 epochs; center, per-client validation loss; right, per-client validation accuracy. Middle row — FL results: left, individual client training losses during federated rounds; center, client validation losses after rounds; right, per-client validation accuracies together with the aggregated FL testset accuracy (magenta). Bottom row — CL (centralized) results: left, pooled training loss; center, pooled validation loss; right, pooled validation accuracy. All loss plots use the same vertical scale for direct visual comparison; accuracy is shown on a 0–1 scale.

*3.2. Local Validation Performance*

Per-client validation metrics computed using client-specific optimized thresholds are reported in Tables 2–4 and visualized in the per-client ROC overlays and confusion matrices (Figure 4). Among the LL models, LL_5 obtained the highest overall scores across most reported metrics on its local validation set (accuracy, AUC, F1 and balanced accuracy), although it did not have the highest sensitivity or specificity. CL, evaluated on each client's local validation set, produced consistently higher performance than the corresponding LL models on the same local splits, and CL's per-client AUCs are consistently high (per-client values appear in Figure 4, middle panel). FL's per-client validation performance lies between the LL ensemble and CL: FL improves over most LL models on average but generally does not reach CL's per-client performance peak.

Statistical testing of AUC values on the local validation sets (Table 6) confirms these trends: CL significantly outperforms LL, and CL also significantly outperforms FL after Bonferroni correction, whereas the difference between FL and LL does not reach statistical significance. The per-client ROC overlays make these relationships visible (Figure 4, left), and the per-client confusion matrices (Figure 4, right) show the class-level error patterns for each paradigm.

**Table 2.** Local Learning models evaluated on their own local validation sets. For each client, the decision threshold is optimized on that client's local training data, and metrics (including Youden's~J) are computed on the corresponding validation set.

| Model | Samples | Opt. thr. | Accuracy | Sensitivity (Recall) | Specificity | F1 (pos) | AUC | Youden's J | Balanced Acc. |
|---|---|---|---|---|---|---|---|---|---|
| Local Learning 0 | 9/56 (pos/neg) | 0.016 | 0.785 | 0.667 | 0.804 | 0.462 | 0.796 | 0.470 | 0.735 |
| Local Learning 1 | 19/46 (pos/neg) | 0.009 | 0.554 | 0.947 | 0.391 | 0.554 | 0.815 | 0.339 | 0.669 |
| Local Learning 2 | 26/36 (pos/neg) | 0.170 | 0.742 | 0.846 | 0.667 | 0.733 | 0.750 | 0.513 | 0.756 |
| Local Learning 3 | 17/48 (pos/neg) | 0.181 | 0.538 | 0.706 | 0.479 | 0.444 | 0.694 | 0.185 | 0.593 |
| Local Learning 4 | 29/50 (pos/neg) | 0.049 | 0.646 | 0.655 | 0.640 | 0.576 | 0.748 | 0.295 | 0.648 |
| Local Learning 5 | 29/33 (pos/neg) | 0.339 | 0.790 | 0.862 | 0.727 | 0.794 | 0.868 | 0.589 | 0.795 |
| Local Learning 6 | 23/53 (pos/neg) | 0.274 | 0.684 | 0.826 | 0.623 | 0.613 | 0.764 | 0.449 | 0.724 |
| Local Learning 7 | 27/41 (pos/neg) | 0.099 | 0.676 | 0.667 | 0.683 | 0.621 | 0.687 | 0.350 | 0.675 |

**Table 3**. Centralized Learning model evaluated on local validation sets. For each client, the decision threshold is optimized on the client's local training data, and metrics are reported on that client's validation set.

| Data | Samples | Opt. thr. | Accuracy | Sensitivity (Recall) | Specificity | F1 (pos) | AUC | Youden's J | Balanced Acc. |
|---|---|---|---|---|---|---|---|---|---|
| Client 0 | 9/56 (pos/neg) | 0.016 | 0.800 | 0.778 | 0.804 | 0.519 | 0.827 | 0.581 | 0.791 |
| Client 1 | 19/46 (pos/neg) | 0.009 | 0.831 | 0.632 | 0.913 | 0.686 | 0.863 | 0.545 | 0.772 |
| Client 2 | 26/36 (pos/neg) | 0.170 | 0.790 | 0.731 | 0.833 | 0.745 | 0.922 | 0.564 | 0.782 |
| Client 3 | 17/48 (pos/neg) | 0.181 | 0.846 | 0.824 | 0.854 | 0.737 | 0.871 | 0.678 | 0.839 |
| Client 4 | 29/50 (pos/neg) | 0.049 | 0.747 | 0.655 | 0.800 | 0.655 | 0.848 | 0.455 | 0.728 |
| Client 5 | 29/33 (pos/neg) | 0.339 | 0.774 | 0.655 | 0.879 | 0.731 | 0.852 | 0.534 | 0.767 |
| Client 6 | 23/53 (pos/neg) | 0.274 | 0.737 | 0.565 | 0.811 | 0.565 | 0.802 | 0.377 | 0.688 |
| Client 7 | 27/41 (pos/neg) | 0.099 | 0.735 | 0.667 | 0.780 | 0.667 | 0.842 | 0.447 | 0.724 |

Table 4. Federated Learning model evaluated on local validation sets. For each client, the decision threshold is optimized on the client's local training data, and metrics are reported on that client's validation set.

| Data | Samples | Opt. thr. | Accuracy | Sensitivity (Recall) | Specificity | F1 (pos) | AUC | Youden's J | Balanced Acc. |
|---|---|---|---|---|---|---|---|---|---|
| Client 0 | 9/56 (pos/neg) | 0.016 | 0.585 | 0.889 | 0.536 | 0.372 | 0.841 | 0.425 | 0.712 |
| Client 1 | 19/46 (pos/neg) | 0.009 | 0.600 | 0.947 | 0.457 | 0.581 | 0.841 | 0.404 | 0.702 |
| Client 2 | 26/36 (pos/neg) | 0.170 | 0.758 | 0.538 | 0.917 | 0.651 | 0.884 | 0.455 | 0.728 |
| Client 3 | 17/48 (pos/neg) | 0.181 | 0.738 | 0.412 | 0.854 | 0.452 | 0.772 | 0.266 | 0.633 |
| Client 4 | 29/50 (pos/neg) | 0.049 | 0.646 | 0.586 | 0.680 | 0.548 | 0.707 | 0.266 | 0.633 |
| Client 5 | 29/33 (pos/neg) | 0.339 | 0.710 | 0.483 | 0.909 | 0.609 | 0.783 | 0.392 | 0.696 |
| Client 6 | 23/53 (pos/neg) | 0.274 | 0.750 | 0.696 | 0.774 | 0.627 | 0.787 | 0.469 | 0.735 |
| Client 7 | 27/41 (pos/neg) | 0.099 | 0.676 | 0.926 | 0.512 | 0.694 | 0.786 | 0.438 | 0.719 |

Table 5. Centralized test performance of Centralized, Federated, and Local Learning models on the centralized test set. For CL and FL, a single global threshold is optimized on pooled training and validation data across all clients; for each LL model, the threshold is optimized on that client's local training and validation data. Youden's~J is computed on the centralized test set at the chosen threshold.

| Model | Samples | Opt. thr. | Accuracy | Sensitivity (Recall) | Specificity | F1 (pos) | AUC | Youden's J | Balanced Acc. |
|---|---|---|---|---|---|---|---|---|---|
| Centralized Learning | 176/369 (pos/neg) | 0.163 | 0.782 | 0.625 | 0.856 | 0.649 | 0.831 | 0.481 | 0.741 |
| Federated Learning | 176/369 (pos/neg) | 0.104 | 0.703 | 0.648 | 0.729 | 0.585 | 0.757 | 0.377 | 0.688 |
| Local Learning 0 | 176/369 (pos/neg) | 0.187 | 0.699 | 0.125 | 0.973 | 0.212 | 0.674 | 0.098 | 0.549 |
| Local Learning 1 | 176/369 (pos/neg) | 0.102 | 0.525 | 0.824 | 0.382 | 0.528 | 0.663 | 0.206 | 0.603 |
| Local Learning 2 | 176/369 (pos/neg) | 0.214 | 0.558 | 0.915 | 0.388 | 0.572 | 0.687 | 0.302 | 0.651 |
| Local Learning 3 | 176/369 (pos/neg) | 0.957 | 0.527 | 0.693 | 0.447 | 0.486 | 0.619 | 0.140 | 0.570 |
| Local Learning 4 | 176/369 (pos/neg) | 0.690 | 0.626 | 0.614 | 0.631 | 0.514 | 0.669 | 0.245 | 0.623 |
| Local Learning 5 | 176/369 (pos/neg) | 0.551 | 0.600 | 0.761 | 0.523 | 0.551 | 0.692 | 0.284 | 0.642 |

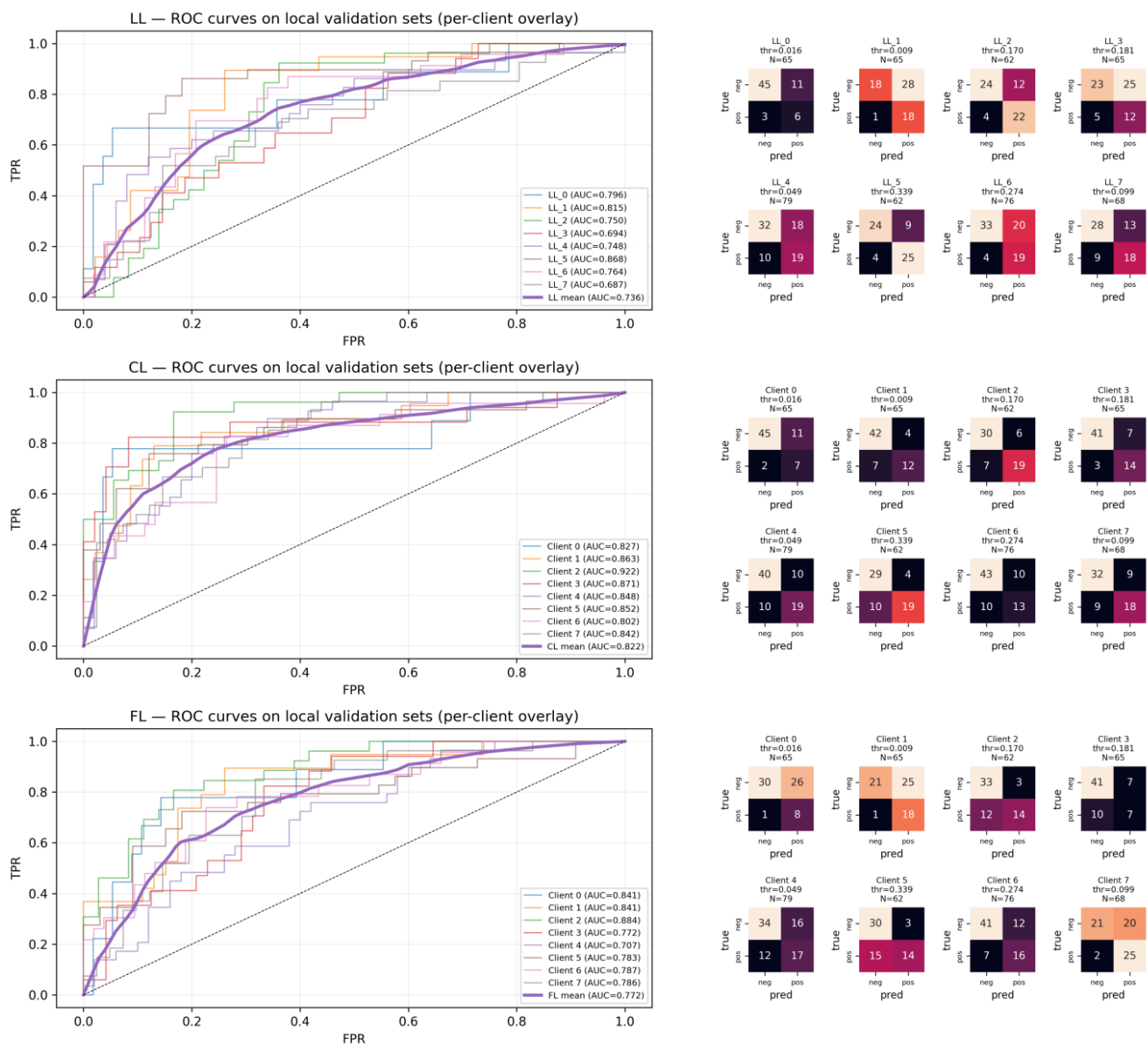

**Figure 4** Performance of all learning paradigms on the local validation sets. For each paradigm (Local Learning, Centralized Learning, Federated Learning), the left panel shows ROC curves for that paradigm evaluated independently on each client's local validation data, using thresholds optimized per client via maximization of Youden's J. The right panel displays the corresponding confusion matrices (one per client), visualizing the distribution of true/false positives and negatives at the individually optimized threshold. This figure highlights how each learning paradigm adapts to heterogeneous client data distributions and how consistently it performs across clients.

*3.3. Centralized Test Performance*

When evaluated on the pooled centralized test set using a single global threshold optimized on pooled training and validation data, CL outperforms the other paradigms on nearly all aggregated metrics. On the centralized test set CL achieves the highest AUC and highest overall accuracy reported (CL AUC = 0.831, CL accuracy ≈ 0.782; see Table 5 and Figure 5). FL is the second-best overall (FL AUC = 0.757, FL accuracy ≈ 0.703), while the LL models perform worst on average. The LL models' centralized test AUCs range across clients (approximately 0.619 to 0.734), with a mean LL AUC of ≈ 0.672 (Figure 5).

DeLong tests on the centralized test set (Table 6) show that CL significantly outperforms FL in the primary comparison. Furthermore, CL significantly outperforms all individual LL models, and FL significantly outperforms seven of the eight LL models after Bonferroni correction; only one LL model (LL_6) does not differ significantly from FL. The centralized ROC overlay (Figure 5, left) clearly orders the curves, CL highest, followed by FL, and then the LL family, while confusion matrices computed under the pooled threshold (Figure 5, right) illustrate that CL yields higher counts of both true positives and true negatives, FL shows intermediate error profiles, and LL models exhibit greater variability across clients.

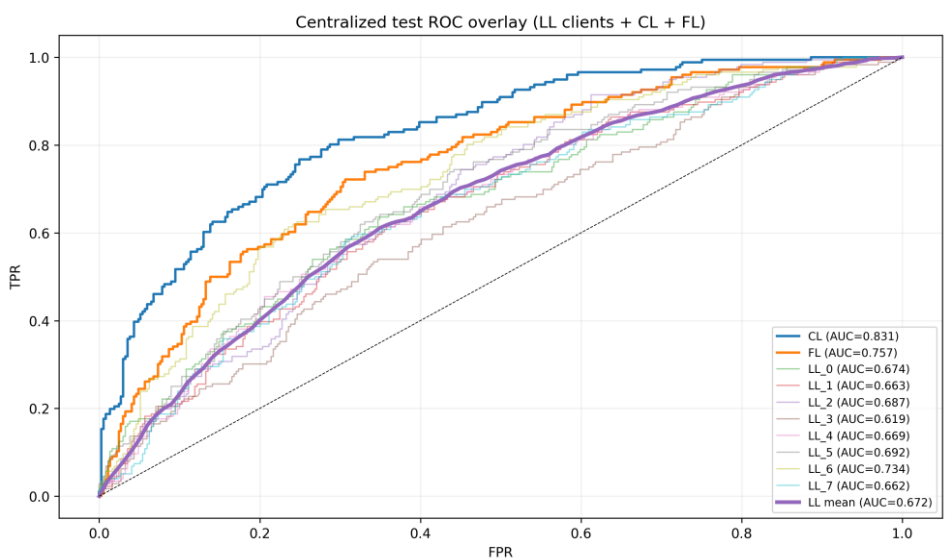
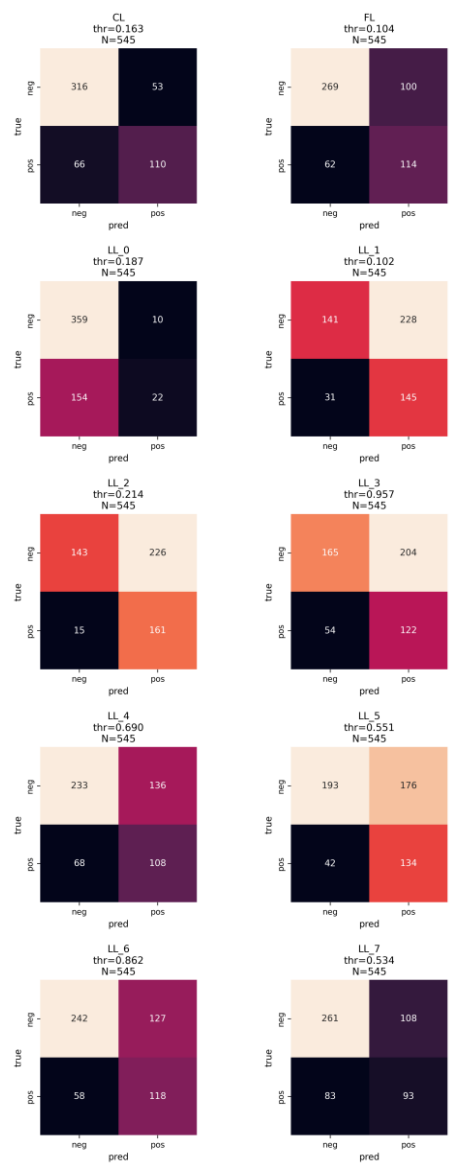

**Figure 5.** Centralized test-set performance across all learning paradigms. Left: ROC curves for the Centralized Learning (CL), Federated Learning (FL), and Local Learning (LL$_{0-7}$) models evaluated on the pooled centralized test set. All models use decision thresholds optimized on the combined centralized validation set to ensure fair comparison. Right: Confusion matrices for each paradigm on the centralized test set, computed using the same pooled-validation threshold. This combined visualization shows both the ranking of models in terms of AUC and their error profiles (false positives/negatives) under a unified and comparable decision rule.

*3.4. Model Explainability*

Grad-CAM overlays for individual test examples (Figure 6) and averaged per-confusion-type heatmaps (Figure 7) highlight the image regions contributing most to each model's decisions. Across paradigms, the stronger models tend to concentrate activation centrally around the third-molar / mandibular-canal region in the image. For the higher-performing CL and FL models the mean Grad-CAMs show well-localized, consistent activation patterns in the areas of clinical interest (Figure 6–7). By contrast, several lower-performing LL models produce more spatially diffuse or inconsistent activations in the mean Grad-CAMs and on misclassified examples (Figure 6–7), reflecting greater heterogeneity in the image regions the models rely on.

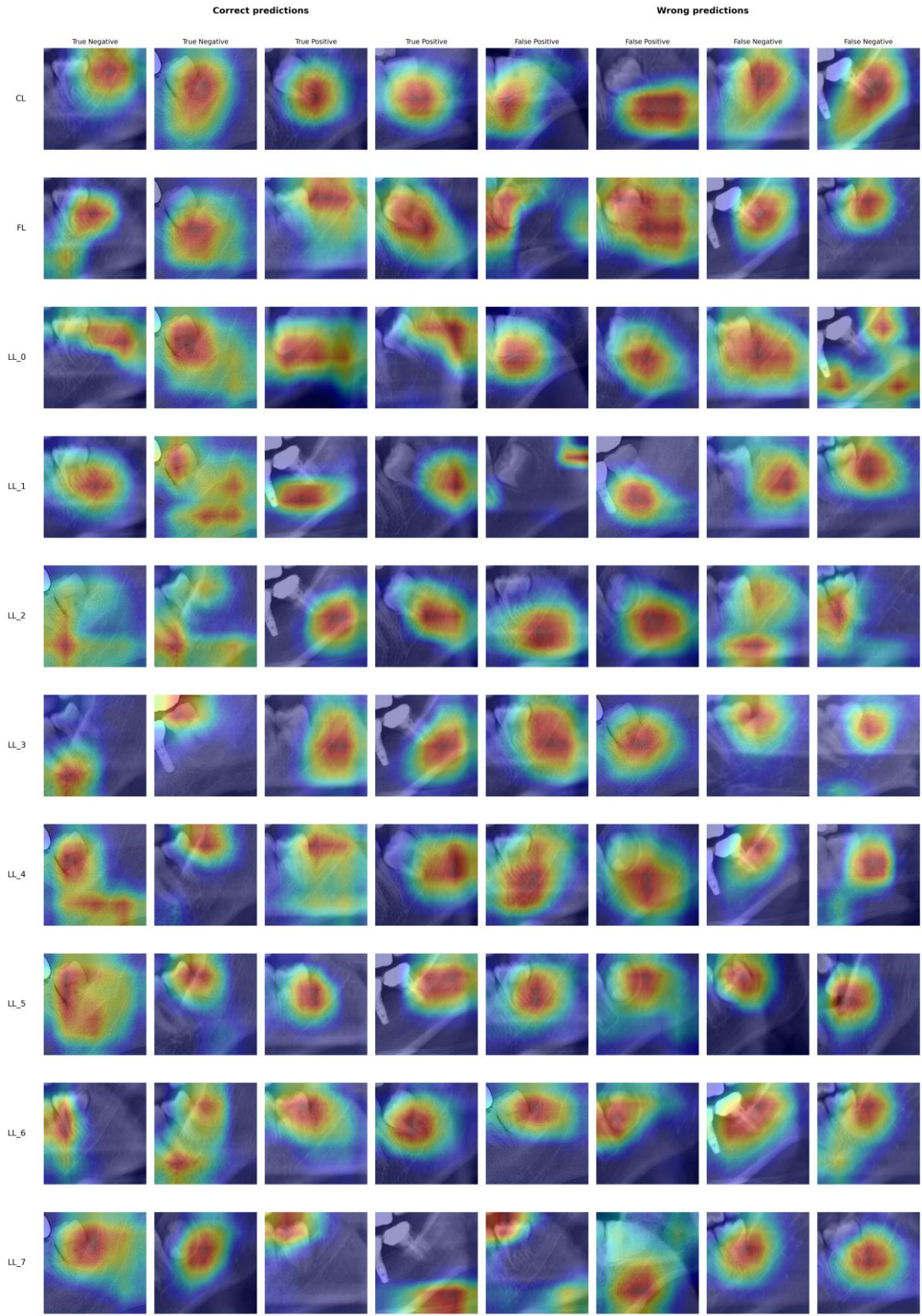

**Figure 6.** Grad-CAM visualizations for the Centralized Learning (CL), Federated Learning (FL), and Local Learning models (LL$_0$–LL$_7$) on the centralized test set. Each row corresponds to one model. The first four columns show correctly classified samples (two true negatives and two true positives), while the last four columns show misclassified samples (two false positives and two false negatives). The heatmaps highlight the image regions that contributed most strongly to each model's decision.

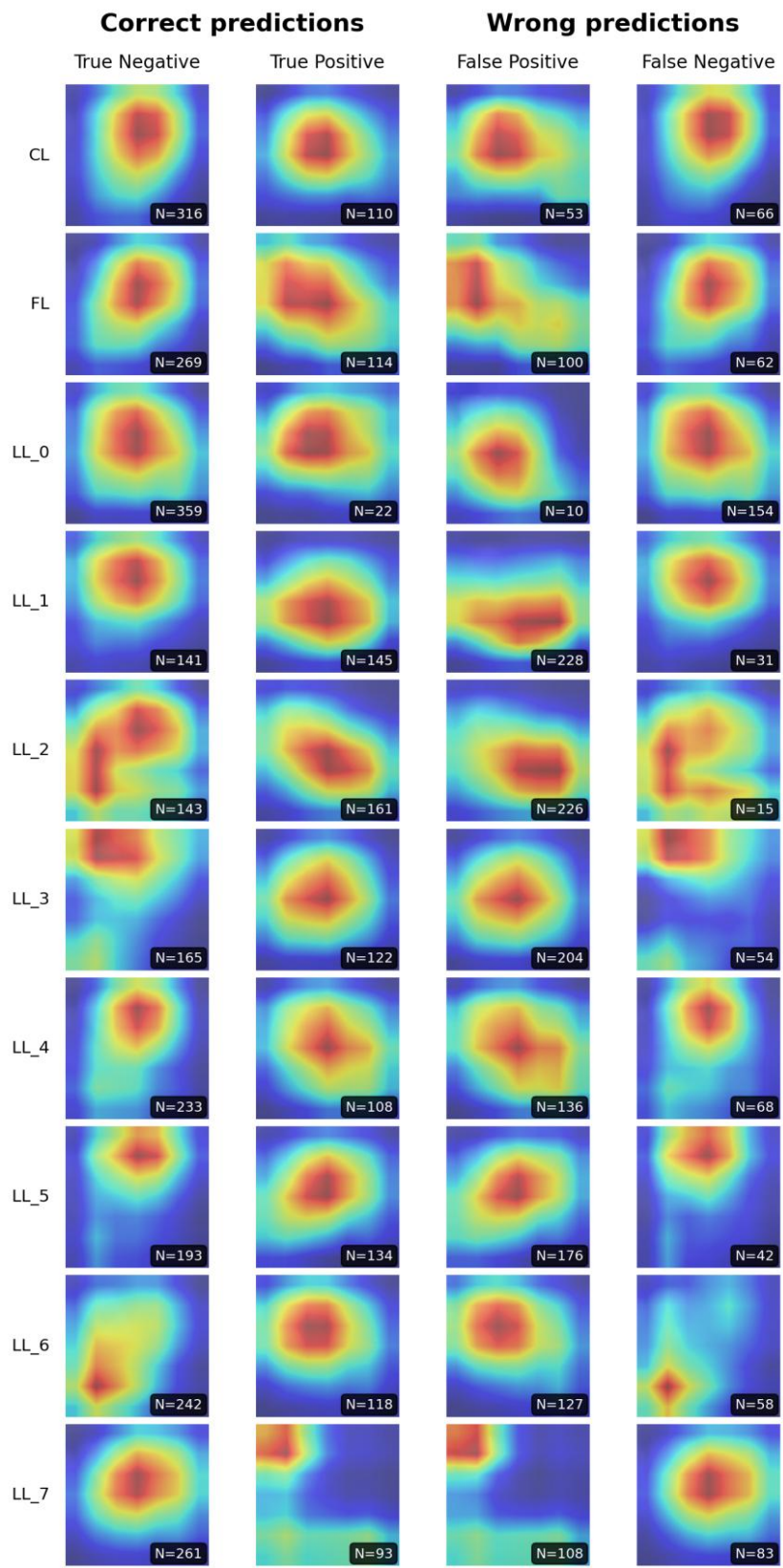

**Figure 7.** Average Grad-CAM heatmaps (overlaid on mean input) for each model on the centralized test set. Rows are models (Centralized Learning — CL, Federated Learning — FL, and Local Learning clients LL_0…LL_7); columns are confusion types computed on test set using optimal thresholds: True Negative, True Positive, False Positive, False Negative. Each panel displays the mean Grad-CAM (jet colormap) over all test samples of that confusion type; the white-on-black label at the lower right reports the number of samples (N) used to form the mean.

## 4. Discussion

The results reported in this work show a clear and consistent pattern: Automatic classification of 3M-MC overlap is feasible to some extent. When data are pooled and centrally available, a conventionally trained centralized model attains the strongest generalization on the pooled test set; federated training (FedAvg-style aggregation) yields a reliable intermediate improvement over purely local training; and individually trained local models attain high within-site performance but fail to generalize across sites. These empirical regularities appear in multiple complementary measurements, i.e., ROC overlays and AUCs, confusion matrices under a common decision rule, training/validation dynamics, and Grad-CAM visualizations, and together provide a coherent story about how model capacity, data heterogeneity and calibration interact in multi-site medical-imaging settings.

*Learning dynamics and overfitting*

The learning curves suggest classical overfitting behavior: training loss falls quickly and stabilizes at low values while validation loss is more unstable and, in many runs, even increases; validation accuracy climbs only slowly and remains noisy. Note that for some clients the minority class was nearly doubled via augmentation; these extra, transformed copies increase apparent training sample size, but are derived from the same sources and can therefore inflate training performance while having a limited or uneven effect on true generalization, which likely contributed to the noisy/slowly improving validation accuracy and occasional rises in validation loss. This divergence indicates that the networks are fitting idiosyncratic patterns in the training splits that do not transfer to held-out data. In practical terms, the curves warn that monitoring training loss alone would have given a falsely optimistic picture; instead, validation loss/accuracy and especially calibration diagnostics are necessary to judge real-world readiness. The phenomenon aligns with broader observations that modern deep networks can be poorly calibrated and often require explicit post-hoc calibration (e.g., temperature scaling and calibration metrics such as ECE and Brier score) [36].

*Why local models fail to generalize*

Importantly, part of the apparent overfitting likely reflects that the 3M-MC relationship is intrinsically difficult to assess on panoramic radiographs. The task relies on subtle and often inconsistent cues (e.g., cortical interruption, canal diversion, root darkening) within a 2D projection of 3D anatomy, where superimposition, distortion, metal artefacts, motion, and exposure differences can obscure relevant structures. Substantial anatomical variation and inter-observer disagreement further introduce label noise. In this context, CNNs may fit superficial or site-specific intensity patterns in the training data rather than robust anatomical features, contributing to the gap between low training loss and unstable validation performance, and ultimately limiting achievable classification performance even under CL. Future work should explore whether this assessment is best treated as a segmentation task rather than a classification task.

Local models exemplify the limitations and risks of narrow, site-specific training. On each client's local validation set, the corresponding LL model typically achieves strong discrimination, but those same models perform poorly on the pooled, centralized test set. Two complementary explanations account for this behavior. First, LL models learn local priorities and imaging artefacts (device-specific contrast, local patient demographics, annotation styles) that are predictive within the originating domain but not universally predictive. Second, the raw score distributions from LL models differ widely between clients: some LL models systematically output low probabilities, others high probabilities.

Consequently, optimal decision thresholds computed per client are extreme and unstable, and when a single pooled threshold is applied these models produce severely skewed false positive / false negative profiles [37]. In short, LL delivers fairly good local accuracy at the cost of poor calibration and poor cross-domain generalization, an unattractive trade-off for applications that require a uniform decision rule or that must perform reliably across institutions. The same is illustrated in ROC overlays: LL models often show high within-client ROC performance, but their centralized ROC curves reveal major cross-domain degradation. This pattern underscores that good local AUCs do not guarantee global generalizability, a model can be well-ranked inside a domain, yet misranked when inputs change.

*Federated learning: compomises and limits*

Federated Learning (FL) reduces many of the extremes seen under LL: averaged models trained across clients are less variable than purely local models, and FL tends to improve aggregated performance relative to an LL ensemble. However, in our experiments FL showed a performance gap to CL. This shortfall likely reflects the combined influence of statistical heterogeneity (non-IID client data) and the limitations of simple averaging (FedAvg) when client updates point in different directions. Alternatives that explicitly address this could be FedProx, FedAdam or FedAdagrad. The validation loss fluctuations observed after aggregation rounds are consistent with known convergence and stability issues under heterogeneity [31, 38]. The empirical picture suggests that FL remains a pragmatic and privacy-preserving compromise and it leverages cross-site information to reduce extreme failure modes, but achieving parity with CL requires algorithmic refinements (heterogeneity-aware optimizers, personalization layers, or small amounts of shared calibration data) [39, 40].

*Explainability evidence*

Explainability visualizations provide supporting qualitative evidence for these quantitative claims. Grad-CAM overlays show that the better performing models (CL and the stronger FL runs) localize attention in anatomically plausible regions near the third molar and mandibular canal; averaged heatmaps for true positives and true negatives tend to be concentrated and consistent [41]. By contrast, many LL models produce more spatially diffuse or inconsistent activations in the averaged maps and on misclassified examples. These saliency patterns are consistent with the idea that models that generalize better learn features anchored to the clinically relevant anatomy, while overfit models latch on to unstable cues. It is worth emphasizing, however, that saliency maps are informational but not definitive explanations: they should be combined with quantitative localization tests (if ground-truth region annotations are available) and subjected to sanity checks to avoid over-interpreting visually persuasive but potentially misleading heatmaps [42].

*Methodological limitations*

Several methodological limitations constrain how broadly the present conclusions can be generalized. Some clients have relatively small sample sizes and imbalanced classes, which magnifies overfitting and makes per-client threshold estimates noisy. While FL would unlock possibility of more clients participating and thereby increase class representation, the data-size increase here was insufficient to produce enough examples of rare classes; an alternative would be for clients to specifically look for underrepresented classes. The FL experiments used a basic FedAvg aggregation; as mentioned above, alternative FL algorithms (FedProx, adaptive aggregation schemes, cluster-based or personalized

FL approaches) have been proposed precisely to mitigate the negative effects of heterogeneity and may substantially reduce the gap to centralized training [39]. Finally, the evaluation protocol here intentionally contrasted locally optimized thresholds (for LL) with a pooled threshold (for centralized test comparisons) in order to highlight calibration mismatches; different operational choices (e.g., always using per-site thresholds in deployment) would change practical outcomes and should be explicitly considered alongside pooled evaluations.

In a realistic federated deployment, one cannot assume access to a pooled global test set or the ability to compute a single "optimal" threshold centrally. A practical, privacy-preserving compromise is for each client to compute summary metrics locally across a range of candidate thresholds, for example Youden's J (or sensitivity/specificity pairs), and send those scalar summaries (not raw data) to the server. The server can then aggregate these J-curves (e.g., average J or robust statistics such as the median or worst-case J) to select a threshold that balances performance across sites. This approach preserves local data privacy while making use of per-site calibration behavior; it also exposes heterogeneity (different prevalence and score bias) that should inform whether a single global rule is appropriate. For this study a pooled global threshold chosen on centralized validation data works well for CL, and it is reasonably appropriate for FL, but it penalizes many LL models whose score distributions are not aligned with the pooled calibration. From a deployment perspective this matters because clinical decision rules are thresholded: a single poorly chosen threshold can either flood clinicians with false alarms or miss true positives, depending on model bias. Therefore, sending calibration metrics as suggested or using other calibration methods such as Platt scaling, Brier scores or temperature scaling is highly recommended [43].

Inspection of the training dynamics further suggests that some clients may behave systematically differently from others in ways that are detectable without access to sensitive data. From a server-side perspective, such differences can be inferred from non-private signals routinely available in federated learning, including local training and validation loss trajectories, validation accuracy trends, update magnitudes, and the consistency of model updates across rounds. For example, a client that reports consistently lower loss and higher validation accuracy than its peers (Client 0 in this scenario, see Figure 3), yet yields strongly asymmetric operational metrics (e.g., very high specificity but poor sensitivity and balanced accuracy), may indicate conservative labeling practices, local shortcut learning, or skewed class distributions (which was the case for this client). Though, in our calibration analysis, Client 0 showed high inter-annotator agreement, suggesting that the unusual validation metrics for that client are unlikely to be explained by labeler idiosyncrasy. By contrast, Client 3 was an outlier in the calibration (lower pairwise weighted κ), and therefore per-client results for Client 3 should be interpreted with caution because annotator variability may contribute to asymmetric operational metrics even though this was not the case. Similar discrepancies have been noted in federated settings where clients differ substantially in data composition or annotation style, and where simple aggregation masks these divergences [31, 38].

Beyond scalar losses and accuracies, servers can monitor update-level statistics that do not reveal raw data, such as the norm of client weight updates, cosine similarity between client updates, or their variance across rounds. Such diagnostics have been proposed to detect stragglers, outliers, or statistically heterogeneous clients and to guide adaptive aggregation strategies without compromising privacy [40, 39].

All these server-side indicators can inform optimization of both training and aggregation. As federated learning moves toward clinical deployment, such non-intrusive monitoring mechanisms are increasingly recognized as essential for identifying hidden heterogeneity, label noise, or systematic bias across sites, and for preventing misleadingly optimistic global metrics that obscure poor real-world performance in specific clients [44].

Lastly, it is important to acknowledge that the quality and consistency of the labeling process may have been significantly influenced by various human and contextual factors. This includes the diverse seniority levels of the labelers, which ranged from recently graduated dentists to experienced academic clinicians. Such inter-rater variability likely contributed to differences in interpretive confidence and diagnostic thresholds. Additionally, labelers operated in heterogeneous environments, each equipped with distinct monitors and lighting setups, which introduced uncontrolled variability in display calibration and ambient conditions. Cognitive fatigue may have also emerged as a critical consideration due to the high volume of radiographs analyzed. Furthermore, decent intra-rater consistency could pose additional challenges that we were unable to quantify, as limited resources prevented us from conducting repeated labeling rounds. Compounding these issues was the inherent ambiguity in class definitions, particularly between classes 1 and 2A. This ambiguity was particularly pronounced in certain instances, where the overlap between the third molar and the inferior alveolar nerve could not be definitively verified due to the poor quality of images within the datasets. Collectively, these factors may have impacted the model's apparent performance, highlighting the need for structured calibration and ongoing reliability assessments in future studies.

## 5. Conclusions

This study empirically compared LL, FL and CL for automatic detection of 3M-MC overlap on cropped panoramic ROIs. Across multiple complementary diagnostics, a consistent pattern was observed: CL attained the strongest generalization on the pooled test set, FL offered a useful privacy-preserving middle ground that improved robustness relative to purely local models, and LL models performed well within-site but generalized poorly across sites. Important practical problems were revealed: pronounced client heterogeneity, score miscalibration, and unstable per-client optimal thresholds, issues that make a single pooled decision rule unreliable for many local models. Server-side monitoring of non-sensitive signals can identify atypical clients and guide aggregation or personalization without sharing sensitive data. Taken together, the findings recommend centralized pooling when feasible, and, when it is not, careful heterogeneity-aware federated strategies, explicit probability calibration, per-site reporting, and non-intrusive server-side monitoring to make federated deployments safer and more reliable. Future work should explore heterogeneity-aware FL algorithms, targeted calibration strategies, and quantitative localization analyses to further close the gap between FL and CL for clinical tasks.

**Supplementary Materials:** No supplementary materials.

**Author Contributions:** Conceptualization, J.R, S.H., S.J. and R.P.; Data Curation, S.H., S.J., M.A., A.C., S.E., F.G., J.I., B.K., R.R. and R.P.; Formal Analysis, J.R., S.H., S.J. and R.P.; Funding acquisition, R.P.; Investigation, J.R., S.H., S.J. and R.P.; Methodology, J.R., S.H., S.J. and R.P.; Resources, S.H., S.J., M.A., A.C., S.E., F.G., J.I., B.K., R.R. and R.P.; Software J.R., S.H., S.J. and R.P.; Supervision, R.P.; Validation, J.R., S.H., S.J. and R.P.; Visualization, J.R., R.P.; Writing – original draft, J.R.; Writing – review & editing J.R., S.H., S.J., M.A., A.C., S.E., F.G., J.I., B.K., R.R. and R.P.

**Funding:** This work was funded by: (1) the Independent Research Fund Denmark, project "Synthetic Dental Radiography using Generative Artificial Intelligence", grant ID 10.46540/3165-00237B, for computational resources; (2) the Aarhus University Centre for Digitalisation, Big Data and Data Analytics and the Circle U Seed Funding Scheme for salary costs.

**Institutional Review Board Statement:** Not applicable.

**Informed Consent Statement:** Not applicable

**Data Availability Statement:** The original data presented in the study are openly available at Kaggle at https://www.kaggle.com/datasets/zwbzwb12341234/a-dual-labeled-dataset, Zenodo at https://zenodo.org/records/7812323#.ZDQE1uxBwUG, Kaggle at https://www.kaggle.com/datasets/humansintheloop/teeth-segmentation-on-dental-x-ray-images, Tufts Dental Database opun request at https://tdd.ece.tufts.edu/ and upon request from authors of 10.1016/j.oooo.2023.12.006 .

**Acknowledgments:** The authors thank the anonymous reviewers for their valuable suggestions. During the preparation of this manuscript/study, the author(s) used ChatGPT, 5.2 for the purposes of language editing, structure refinement and clarity. All scientific content, hypotheses and methods were developed by the authors. The authors have reviewed and edited the output and take full responsibility for the content of this publication.

**Conflicts of Interest:** The authors declare no conflicts of interest.

# Abbreviations

The following abbreviations are used in this manuscript:

| | |
|---|---|
| LL | Local Learning |
| CL | Centralized Learning |
| FL | Federated Learning |
| AUC | Area under curve |
| CBCT | Cone beam computed tomography |
| ROI | Region of interest |
| IID | Independent and identically distributed |
| AI | Artificial Intelligence |
| ROC | Receiver operating characteristic |

# Appendix A

*Appendix A.1 – Significance Testing*

**Table 6.** Statistical significance testing of AUC differences between learning paradigms. Local validation performance compares discrimination ability across clients using paired tests, while centralized test performance evaluates global generalization on a held-out test set. Multiple-testing correction is applied within each family of related hypotheses as described in the table footnotes

| Evaluation setting | Comparison | Test | Statistic | p-value | Significant |
|---|---|---|---|---|---|
| **Local validation sets (AUC, paired across clients)** | | | | | |
| Local validation | CL vs LL | Wilcoxon | – | 0.0156 | YES* |
| Local validation | FL vs LL | Wilcoxon | – | 0.2500 | NO |
| Local validation | CL vs FL | Wilcoxon | – | 0.0156 | YES* |
| **Centralized test set (AUC, pooled)** | | | | | |
| Centralized test | CL vs FL | DeLong | z = −3.28 | 1.043e−03 | YES† |
| Centralized test | CL vs LL_0 | DeLong | z = −6.63 | 3.380e−11 | YES‡ |
| Centralized test | CL vs LL_1 | DeLong | z = −7.37 | 1.730e−13 | YES‡ |
| Centralized test | CL vs LL_2 | DeLong | z = −5.98 | 2.290e−09 | YES‡ |
| Centralized test | CL vs LL_3 | DeLong | z = −7.70 | 1.407e−14 | YES‡ |
| Centralized test | CL vs LL_4 | DeLong | z = −5.97 | 2.398e−09 | YES‡ |
| Centralized test | CL vs LL_5 | DeLong | z = −5.79 | 7.040e−09 | YES‡ |
| Centralized test | CL vs LL_6 | DeLong | z = −3.97 | 7.124e−05 | YES‡ |
| Centralized test | CL vs LL_7 | DeLong | z = −6.58 | 4.710e−11 | YES‡ |
| Centralized test | FL vs LL_0 | DeLong | z = −3.35 | 8.203e−04 | YES‡ |
| Centralized test | FL vs LL_1 | DeLong | z = −4.04 | 5.269e−05 | YES‡ |
| Centralized test | FL vs LL_2 | DeLong | z = −3.19 | 1.442e−03 | YES‡ |
| Centralized test | FL vs LL_3 | DeLong | z = −6.52 | 7.172e−11 | YES‡ |
| Centralized test | FL vs LL_4 | DeLong | z = −4.46 | 8.378e−06 | YES‡ |
| Centralized test | FL vs LL_5 | DeLong | z = −2.93 | 3.432e−03 | YES‡ |
| Centralized test | FL vs LL_6 | DeLong | z = −1.03 | 3.033e−01 | NO |
| Centralized test | FL vs LL_7 | DeLong | z = −3.99 | 6.631e−05 | YES‡ |

\* Local validation: paired Wilcoxon signed-rank tests with Bonferroni correction over three planned comparisons ($\alpha=0.05/3$).

† Centralized test (primary analysis): DeLong's test for correlated ROC curves, uncorrected ($\alpha=0.05$).

‡ Centralized test (robustness analyses): DeLong's test with Bonferroni correction over eight Local Learning models per paradigm ($\alpha=0.05/8$).

*Appendix A.2 – Inter-annotator reliability*

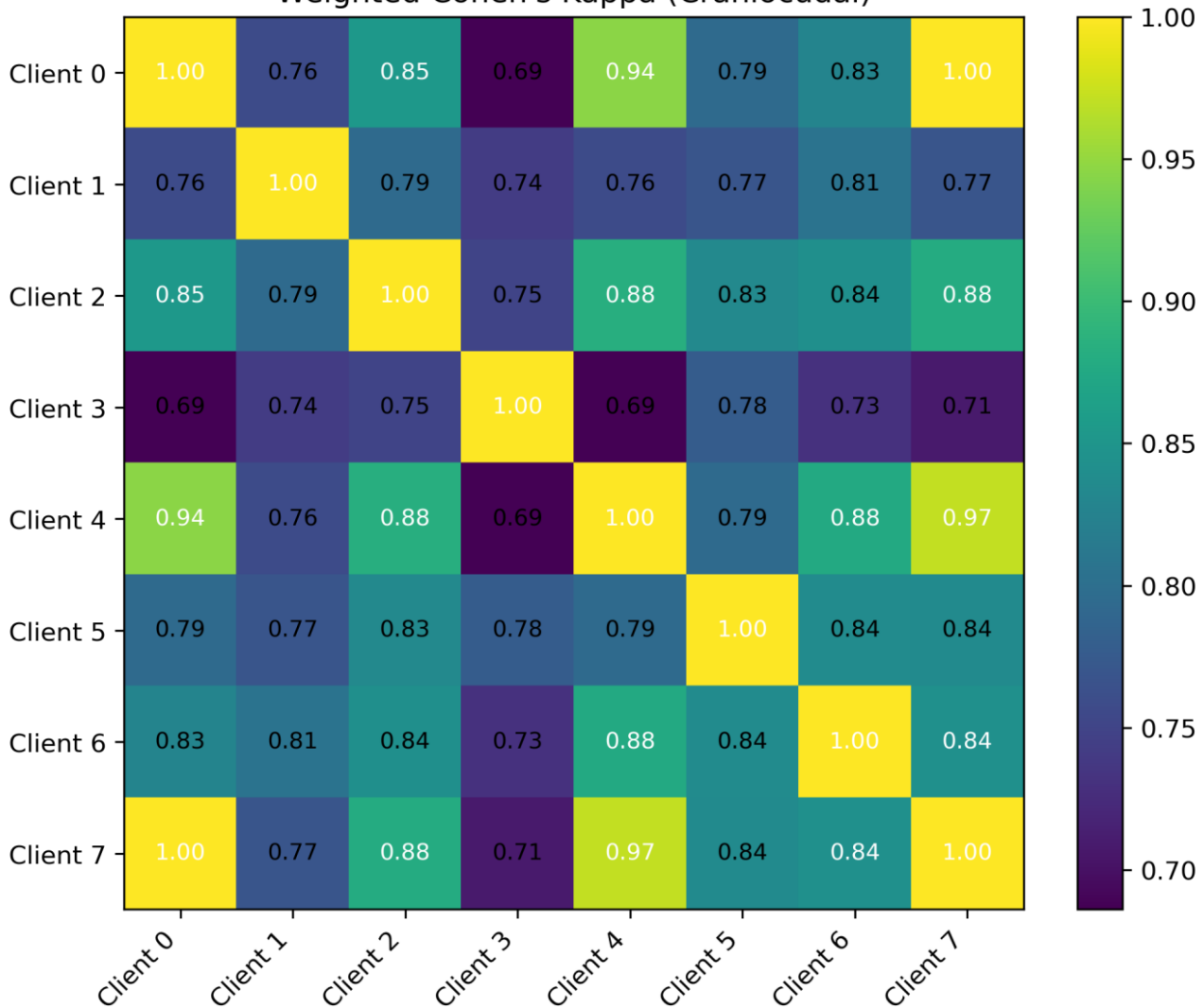

**Figure 8.** Weighted Cohen's κ (craniocaudal). Heatmap showing pairwise quadratic-weighted Cohen's κ between clients for the craniocaudal classification task. Higher values indicate stronger agreement. Diagonal values equal 1.00 by definition.